\documentclass[12pt,aps, pra, superscriptaddress]{revtex4-2}

\usepackage{amsmath, amssymb, mathtools, amsthm}
\usepackage[colorlinks=true,linkcolor=blue]{hyperref}
\usepackage{graphicx}

\newcommand{\tr}{{\text{tr}}}

\newcommand{\T}[1]{#1}

\numberwithin{equation}{section}



\begin{document}

\title{A simple realization of Weyl--Heisenberg covariant measurements}

\author{Sachin Gupta}
\affiliation{QBism Group, Physics Department, University of Massachusetts Boston, Boston, MA, USA}

\author{Matthew B. Weiss}
\affiliation{QBism Group, Physics Department, University of Massachusetts Boston, Boston, MA, USA}

\begin{abstract}
Informationally complete (IC) measurements are fundamental tools in quantum information processing, yet their physical implementation remains challenging. By the Naimark extension theorem, an IC measurement may be realized by a von Neumann measurement on an extended system after a suitable interaction. In this work, we elaborate on a simple algorithm for realizing Naimark extensions for rank-one Weyl--Heisenberg covariant informationally complete measurements in arbitrary finite dimensions. Exploiting Weyl--Heisenberg covariance, we show that the problem reduces to determining a $d \times d$ unitary from which the full $d^2 \times d^2$ unitary interaction can be constructed. The latter unitary enjoys a block-circulant structure which allows e.g., for an elegant optical implementation. We illustrate the procedure with explicit calculations for qubit, qutrit, and ququart SIC-POVMs. Finally, we show that from another point of view, this method amounts to preparing an ancilla system according to a so-called fiducial state, followed by a generalized Bell-basis measurement on the system and ancilla. These results provide a straightforward framework for implementing informationally complete measurements in the laboratory suitable for both qubit and qudit based systems.
\end{abstract}


\maketitle


\section{Introduction}

The efficient implementation of informationally complete measurements (IC-POVMs) remains one of the central challenges in quantum information processing and quantum computation. Remarkably, assigning probabilities to the outcomes of an informationally complete measurement is equivalent to assigning a density matrix, or quantum state, making IC-POVMs an invaluable tool both theoretically and experimentally \cite{DAriano2004,Scott_2006, fuchs2019qbism}. One common way to realize such a measurement is via a so-called Naimark extension. It is a theorem that \emph{any} quantum measurement can be realized by adjoining an ancillary system, instigating an interaction between the system of interest and the ancilla, subsequently performing a standard von Neumann measurement on the ancilla as well as potentially on the system itself \cite{kraus1983states}. That said, not only is there a great deal of freedom in the unitary interaction that realizes a particular measurement, but it is not always practical to implement such arbitrary interactions in the laboratory. The situation becomes simpler when the measurement is group-covariant \cite{DAriano2004}. In this work, we explore a particularly straightforward method for realizing a Naimark extension for any rank-one informationally complete measurement covariant under the discrete Weyl--Heisenberg (WH) group in arbitrary dimension.  Indeed, for generic states, such measurements are informationally complete. This construction is timely given the growing interest in higher-dimensional quantum computing platforms, for example, higher-level trapped-ion qudits \cite{IonTrapHigherDimensional2025} and multilevel superconducting circuits \cite{Goss2024}. 

The Naimark extension we discuss draws on two threads. On the one hand, Tabia \cite{Tabia2012} proposes an elegant optical scheme to realize the special case of symmetric informationally complete (SIC) POVMs in dimensions $d=2$ and $3$ using multiport interferometric devices. These setups demonstrate that SIC-POVMs can be achieved through compact optical circuits that effectively encode the Naimark extension. In particular, Tabia chooses the Naimark unitary $U$ to be block circulant so that the implementation can exploit its block diagonalization. Motivated by this approach, we show how his method can be extended in a natural way to arbitrary dimension by imposing three symmetry assumptions on $U$. In fact, any freedom in $U$ (a $d^2 \times d^2$ matrix) thereby reduces to the choice of a $d \times d$ unitary, one of whose rows is fixed. On the other hand, in their influential work on group symmetry and informational completeness \cite{DAriano2004}, D'Ariano \textit{et al.} show that one can realize WH covariant IC-POVMs by performing a generalized Bell measurement on the system and an ancilla prepared in the so-called WH ``fiducial'' state.  The generalized Bell-basis was already used in the original quantum teleportation paper \cite{Bennett1993} and further explored in \cite{Werner2001}; in \cite{DAriano2000} they were placed in a larger group-theoretic context; \cite{DAriano2004_2} used them to construct so-called ``universal detectors,'' while \cite{Popp2024} provides a more recent study of them. Generalized Bell-basis implementations of SIC-POVMs in particular appear more recently, for example, \cite{Jiang2020} and in \cite{You2025}  (in the single qubit case). These procedures may be compared with the general group theoretic methods of \cite{Janzing2006, decker2005implementation}. We show that in fact our extension of Tabia's method realizes precisely a generalized Bell-basis implementation of the WH-POVM. Moreover, this analysis makes clear that the three assumptions that guide the block-matrix construction of $U$ are equivalent to the single assumption that, by appropriately choosing the ancilla state, one can realize $d$ different WH-POVMs with orthogonal fiducial states. One application, therefore, is to the compact implementation of so-called compound SICs \cite{SICQuantumKey2020}. Finally, we note that whereas the block-matrix formalism of Section~\ref{from blocks} lends itself to optical implementations where one assigns basis vectors in $\mathcal{H}_{d^2}$ to optical paths that can be individually addressed, the generalized Bell-basis implementation of Section~\ref{gen_bell_img} is more appropriate for multiqudit platforms where one treats $\mathcal{H}_{d^2} \simeq \mathcal{H}_d \otimes \mathcal{H}_d$. In the latter context, the transformation to the generalized Bell-basis can be implemented by a qudit controlled-shift operator followed by a qudit Fourier transform \cite{Karimipour2002, King2024, mukherjee2025certifyingdimensionalityquantumchannel}. In Appendix \ref{qubits}, we explain how when $d=2^n$, these qudit operations can be broken down into one-and two-qubit operations for implementation on qubit-based quantum computers.

The remainder of this paper is organized as follows. In Section~\ref{preliminaries}, we review the formalism of Weyl--Heisenberg covariant measurements. In Section~\ref{from blocks}, we develop a Naimark extension for any rank-one WH covariant measurement from a block--matrix perspective, showing how three natural assumptions lead to a convenient block diagonalization appropriate for optical implementations. Concrete examples in dimensions $d=2, 3$, and $4$ are discussed in Section~\ref{examples}, in particular, in $d=2,4$ our examples correspond to compound SICs \cite{SICQuantumKey2020}. In Section~\ref{gen_bell_img}, we show how the previous method is equivalent to a generalized Bell-basis implementation of the WH-POVM, and comment on how this may be implemented on multiqudit and multiqubit systems. We conclude in Section~\ref{conclusion} with further remarks on the experimental implementation of WH-POVMs and suggestions for lines of future research.

\section{Weyl--Heisenberg covariant rank-one measurements}
\label{preliminaries}
A quantum measurement with a finite number $n$ of outcomes may be modeled by a collection of operators $\{E_j\}_{j=1}^n$ that act on a $d$-dimensional Hilbert space $\mathcal{H}_d$ and satisfy
\begin{align}
    \forall j: E_j \ge 0, \qquad \sum_j E_j = I_d.
\end{align}
If each element $E_j$ is unbiased (with equal trace) and rank-one, we may write $E_j = \frac{1}{d}|\psi_j\rangle\langle \psi_j|$, where $\{|\psi_j\rangle\}$ are normalized vectors. When the set $\{E_j\}$ spans the $d^2$-dimensional space of operators on $\mathcal{H}_d$, the measurement is called \emph{informationally complete} (IC). Such measurements separate quantum states: no two distinct states are assigned the same outcome probabilities under an IC measurement. When $n=d^2$, we call the measurement \emph{minimal} informationally complete (MIC): in this case, the $\{E_i\}$ must in fact be linearly independent, and this is the case we will focus on.

A particularly important and well-studied class of minimal informationally complete measurements consist of those covariant under the qudit Weyl--Heisenberg group \cite{Weyl1950,schwinger1960unitary}. The qudit WH group acts irreducibly on $\mathcal{H}_d$ through two unitary operators $X$ and $Z$, known as the shift and the clock operators, defined by
\begin{equation}
    X|k\rangle = |k+1\rangle, \qquad Z|k\rangle = \omega^{k} |k\rangle,
\end{equation}
where $\omega = e^{2\pi i / d}$, $\{|k\rangle\}$ is an orthonormal basis which we take to be the computational basis, and index arithmetic is understood mod $d$. Up to phase, the group elements are then given by
\begin{equation}
    D_{jk} = X^{j} Z^{k}, \qquad j,k \in  \mathbb{Z}_d.
\end{equation}
The $\{D_{jk}\}$ form an orthonormal unitary operator basis as $\tr(D_{jk}^\dagger D_{j^\prime k^\prime })=d\delta_{j,j^\prime}\delta_{k, k^\prime}$. We further note that $X$ and $Z$ are Fourier conjugates: letting $F=\frac{1}{\sqrt{d}}\sum_{jk}\omega^{jk}|j\rangle\langle k|$, we have $X = F^\dagger Z F$. 

Given a single normalized \emph{fiducial state} $|\phi\rangle$, the WH group generates an orbit of $d^2$ vectors
\begin{equation}
    |\phi_{jk}\rangle = D_{jk} |\phi\rangle,
\end{equation}
such that if we let $E_{jk} = \frac{1}{d} |\phi_{jk}\rangle \langle \phi_{jk}|$, by Schur's lemma, $\sum_{jk} E_{jk}=I$. The resulting measurement is WH covariant. Moreover, as long as $\forall j, k: \langle \phi | D_{jk}^\dagger |\phi\rangle \neq 0$, it will be informationally complete \cite{DAriano2004}. In particular, all known symmetric informationally complete (SIC) measurements are covariant under the qudit Weyl--Heisenberg group \cite{fuchs2017sic,appleby2017sics,appleby2016generating,Scott_2006} (with the sole exception of the Hoggar SIC \cite{stacey2021first}). A set of SIC states satisfies
\begin{equation}
    |\langle \phi_{jk}| \phi_{j^\prime k^\prime} \rangle|^2 = \frac{d\delta_{j,j^\prime}\delta_{k,k^\prime}+ 1}{d+1},
\end{equation}
implying that $\{|\phi_{jk}\rangle\langle \phi_{jk}|\}$ form a regular simplex inscribed in the space of pure quantum states. SICs offer a variety of practical advantages in many fields, from entanglement detection \cite{SICEntanglement2018} and quantum key distribution \cite{SICQuantumKey2020} to optimal quantum-state tomography \cite{Scott_2006, ZhuTomoography2011, PETZ2012161}. They have deep connections to algebraic number theory \cite{appleby2016generating, appleby2017sics, bengtsson2017number,appleby2018constructing}, higher-dimensional sphere packing \cite{stacey2017sporadic}, Lie and Jordan algebras \cite{appleby2011lie, appleby2013group}, finite groups \cite{zhu2015super, stacey2016geometric}, and foundational issues in quantum mechanics \cite{fuchs2010qbism, tabia2012experimental, Fuchs2013BayesianCoherence, huangjun2012quantum, graydon2016quantum, Zhu2016Quasiprobability, Szymusiak2026InformationalPower, appleby2017Qplex, fuchs2019qbism, stacey2016geometric, appleby2013group}.  In particular, WH covariant SIC fiducial states have maximal magic \cite{cuffaro2024quantumstatesmaximalmagic}.

\section{A Symmetry-Driven Block Construction}
\label{from blocks}
In this section, we outline a block-matrix oriented algorithm for constructing a Naimark extension for any rank-one WH covariant measurement in arbitrary dimension. The Naimark extension theorem \cite{kraus1983states} states that any quantum measurement $\{E_i\}$ on a $d$-dimensional quantum system can be realized by a von Neumann measurement on a larger Hilbert space. In the simplest case, one may embed an input state $|\psi\rangle$ on $\mathcal{H}_d$ as a state $|\psi^\prime\rangle$ on a larger Hilbert space $\mathcal{H}_n$, where $n$ is the number of measurement outcomes. The theorem guarantees the existence of a unitary $U$ on $\mathcal{H}_n$ such that
\begin{align}
P(i) = \langle \psi|E_i|\psi\rangle = |\langle i|U|\psi'\rangle|^2.
\end{align}
Such a unitary, however, is not unique, and finding one that lends itself to experimental implementation is generally not a trivial task. 

As we are interested in performing a Weyl--Heisenberg covariant measurement (with $d^2$ outcomes), we may formally identify the larger Hilbert space with the tensor product of two $d$-dimensional Hilbert spaces, $\mathcal{H}_n \simeq \mathcal{H}_d\otimes \mathcal{H}_d$, and embed the input state as e.g., $|\psi\rangle \otimes |0\rangle = |\psi,0\rangle$, where $|0\rangle$ denotes the first computational basis state. In particular, if $|\psi\rangle = \sum_i \psi_i |i\rangle$, we can write the embedded state as
\begin{align}
|\psi, 0\rangle = \begin{pmatrix} \psi_0 \\ 0 \\ \vdots \\ 0 \\ \psi_1 \\ 0 \\ \vdots \\ 0 \\ \psi_{d-1} \\ 0 \\ \vdots \\ 0 \end{pmatrix},
\end{align}
where a component appears every $d$ entries. Likewise, writing the desired fiducial $|\phi\rangle = \sum_i \phi_i |i\rangle$, we seek a unitary of the form

\begin{align}
\label{blockU}
U &= \frac{1}{\sqrt{d}}
\left(\begin{array}{cc|cc|cc|cc} 
\phi^*_0 & \cdots & \phi^*_1 & \cdots & \cdots & \cdots & \phi^*_{d-1} & \cdots \\
\phi^*_0 & \cdots & (\omega \phi_1)^* & \cdots & \cdots & \cdots & (\omega^{d-1}\phi_{d-1})^* & \cdots \\
\vdots & \vdots & \vdots & \vdots & \vdots & \vdots & \vdots & \vdots \\
\phi^*_0 & \cdots & (\omega^{d-1}\phi_1)^* & \cdots & \cdots & \cdots & (\omega \phi_{d-1})^* & \cdots \\
\hline 
\phi^*_{d-1}& \cdots & \phi^*_0 & \cdots & \cdots & \cdots & \phi^*_{d-2} & \cdots \\
(\omega^{d-1}\phi_{d-1})^* & \cdots & \phi_0^* & \cdots & \cdots & \cdots & (\omega^{d-2}\phi_{d-2})^* & \cdots \\
\vdots & \vdots & \vdots & \vdots & \vdots & \vdots & \vdots & \vdots \\
(\omega \phi_{d-1})^* & \cdots & \phi^*_0 & \cdots & \cdots & \cdots & (\omega^2  \phi_{d-2})^* & \cdots \\
\hline 
\vdots & \vdots &\vdots & \vdots & \vdots & \vdots & \vdots & \vdots \\
\hline 
\phi^*_{1}& \cdots & \phi^*_2 & \cdots & \cdots & \cdots & \phi^*_{0} & \cdots \\
(\omega \phi_{1})^* & \cdots & (\omega^2 \phi_2)^* & \cdots & \cdots & \cdots & \phi^*_{0} & \cdots \\
\vdots & \vdots & \vdots & \vdots & \vdots & \vdots & \vdots & \vdots \\
(\omega^{d-1} \phi_{1})^* & \cdots & (\omega^{d-2}\phi_2)^* & \cdots & \cdots & \cdots &  \phi^*_{0} & \cdots
\end{array}\right).
\end{align}
Considering the filled-in entries only, the first $d$ rows of this matrix correspond to the action of the operators $\{D_{0k}^\dagger = Z^{-k}\}_{k=0}^{d-1}$ on the Hermitian conjugate of the fiducial ($\langle \phi|$), with the components arranged every $d$ entries: in other words, $\langle \phi|D_{0k}^\dagger \otimes \langle 0|$. The next $d$ rows arise from the action of $D_{1k}^\dagger$ on $\langle \phi|$: the remaining rows are filled analogously. Therefore, when this matrix acts on the embedded input state $|\psi, 0\rangle$, we have $(\langle \phi | D_{jk}^\dagger  \otimes \langle 0|)|\psi, 0\rangle = \langle \phi | D_{jk}^\dagger |\psi\rangle=\langle \phi_{j,k}|\psi\rangle$, yielding the proper amplitudes for a subsequent computational basis measurement on the extended system to realize the  WH covariant measurement with the desired fiducial $|\phi\rangle$. It remains to fill in the $(d^4-d^3)$ missing entries so that $U$ is unitary in as systematic a manner as possible.

We begin by observing that because of our choice of embedding, the filled-in entries of $U$ have a block circulant structure. We wager that we can choose the rest of the entries of $U$ to have the same structure, namely
\begin{align}
    U =
    \begin{pmatrix}
    S_0 & S_1 & \cdots & S_{d-1} \\
    S_{d-1} & S_{0} & \cdots & S_{d-2} \\
    \vdots & \vdots & \ddots & \vdots \\
    S_{1} & S_{2} & \cdots & S_{0}
    \end{pmatrix}, && S_{k} =
    \frac{1}{\sqrt{d}}\begin{pmatrix}
         \phi_k^* & \cdots\\
         (\omega^{k}\phi_{k})^* & \cdots \\
         \vdots & \vdots \\
         (\omega^{k(d-1)} \phi_{k})^* & \cdots
    \end{pmatrix},
\end{align}
where each $S_k$ is a $d\times d$ matrix. The constraint that the matrix $U$ is unitary, i.e., $U^\dagger U = I$, translates into $d$ equalities which must be satisfied, 
\begin{equation}
    \label{block_constraints}
   \forall k: \sum_{j}S_{j}^\dagger S_{j+k}  = \delta_{k,0}I.
\end{equation}
Noting that the first column of each $S_k$ is just a scalar multiple of a column of the inverse Fourier matrix $F^\dagger = \frac{1}{\sqrt{d}} \sum_{jk}\omega^{-jk}|j\rangle\langle k|$, let us add to our block circulant assumption the ansatz that each block $S_k$ may be expressed as the outer product of a column $|\tilde{f}_k\rangle=F^\dagger |k\rangle$ of $F^\dagger$ and a (transposed) column $\langle \tilde{m}_j| = \langle j|M^T$ of a matrix $M$ whose first row is the fiducial $\langle \phi|$, that is, 
\begin{align}
 S_j = |\tilde{f}_j\rangle\langle \tilde{m}_j|.
\end{align}
Because $F^\dagger$ is unitary, the $\{ |\tilde{f}_j\rangle \}$ form an orthonormal basis. Consequently, reconsidering Eq.~(\ref{block_constraints}), we have
\begin{align}
\forall k: \sum_j S_j^\dagger S_{j+k} = \sum_j |\tilde{m}_j \rangle\langle \tilde{f}_j|\tilde{f}_{j+k}\rangle\langle \tilde{m}_{j+k}| = \sum_j \delta_{j, j+k} |\tilde{m}_j\rangle\langle \tilde{m}_{j+k}| = \delta_{k,0}I.
\end{align}
On the one hand, when $k=0$, we have
\begin{align}
I = \sum_j |\tilde{m}_j\rangle\langle \tilde{m}_j| = \sum_j M^* |j\rangle\langle j|M^T = M^*M^T \Longrightarrow M M^\dagger = I,
\end{align}
which implies that $M$ is unitary, its rows and columns each forming an orthonormal basis. Meanwhile, for $k\neq 0$, the constraint is automatically satisfied by the orthogonality of the columns $|\tilde{f}_j\rangle$. 

We conclude that the following three assumptions are sufficient to fully determine $U$: a) $U$ has a block circulant structure; b) that the blocks may be expressed as outer products $S_j = |\tilde{f}_j\rangle\langle \tilde{m}_j|$; and c) that $\langle \tilde{m}_j| = \langle j|M^T$ for a unitary matrix $M$ whose first row is the fiducial $\langle \phi|$. The problem of finding a $d^2 \times d^2$ unitary matrix $U$ thereby reduces to the problem of finding a $d \times d$ unitary matrix $M$, providing a quadratic reduction in complexity. Since the first row of $M$ corresponds to (the Hermitian conjugate) of the desired fiducial state, it suffices to determine $(d-1)$ orthonormal vectors to complete the matrix $M$.

\begin{figure}[t!]
    \includegraphics[width=1.0\linewidth]{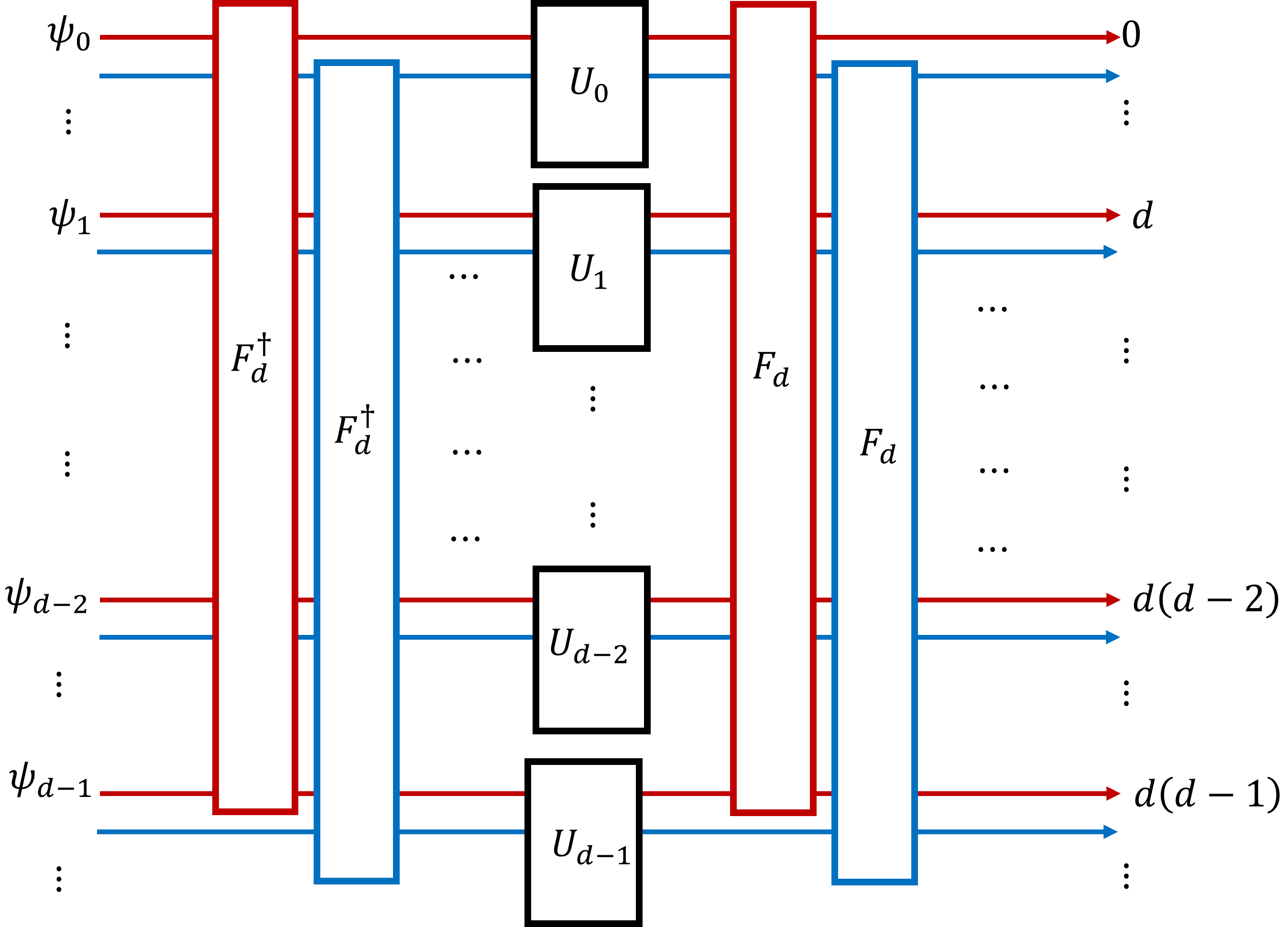}
    \caption{Schematic diagram illustrating the implementation of a rank-one Weyl--Heisenberg--covariant POVM in arbitrary dimension~$d$. The input state $|\psi\rangle$ is injected through the modes $(0,\, d,\, \ldots,\, d(d-1))$. The Fourier transform $F_d$ and its inverse $F_d^\dagger$ act on the modes indicated by the color coding in the diagram. The unitary $U_0$ operates on modes $(0,1,\ldots,d-1)$, $U_1$ on $(d,d+1,\ldots,2d-1)$, and, more generally, $U_{k}$ acts on $(kd, kd+1,\ldots,(k+1)d-1)$ for $k=0,\ldots,d-1$. The detection events at the numbered output ports correspond to the outcomes of the measurement.}
    \label{fig: 1}
\end{figure}

We now observe that the symmetries of this Naimark extension imply a compact, efficient implementation. It is well known that circulant matrices are diagonalized by the Fourier transform \cite{Gray2005}: since $U$ is block circulant, it may be block diagonalized,
\begin{align}
\label{block_diag}
U = (F^\dagger \otimes I)\begin{pmatrix}U_0 & & & \\ & U_1 & & & \\ & & \ddots & \\ & & & U_{d-1}\end{pmatrix}(F \otimes I).
\end{align}
A circulant matrix $C$ has eigenvalues $\lambda_j = \sum_k \omega^{-jk} c_k$, where $c_k$ denotes the component in the first row, $k$th column of $C$. Analogously, the blocks in the block decomposition of $U$ can be expressed as $U_j = \sum_k \omega^{-jk} S_k$. Using the form of $S_k$, we find
\begin{align}
U_j &= \sum_k \omega^{-jk} |\tilde{f}_k\rangle\langle \tilde{m}_k| = \sum_k \omega^{-jk} F^\dagger |k\rangle\langle k | M^T = F^\dagger Z^{-j} M^T.
\end{align}
Altogether this suggests (as in \cite{Tabia2012}) the elegant optical implementation depicted schematically in Figure~\ref{fig: 1}. A single photon is prepared according to a state $|\psi\rangle$, expressing a superposition over $d$ different optical paths. There are $d^2$ total paths. A combination of mirrors, phase shifters, and beam splitters implement the Fourier transform $F$, its Hermitian conjugate, as well as powers of $Z$ and $M^T$, acting on selected subsets of modes. For example, the unitary matrix $U_0$ acts on the first $d$ input modes, $U_1$ on the next $d$, and so on. The Fourier transform acts on every $d$th input: specifically, on the inputs labeled by $(0,\, d,\, 2d,\, \ldots,\, (d-1)d)$, then on inputs $(1,\, d+1,\, 2d+1,\, \ldots,\, (d-1)d+1)$, and so forth. Finally, the photon is detected along one of the $d^2$ paths: this realizes the WH covariant measurement. Its implementation thereby reduces to determining an explicit realization of the smaller unitary matrix $M^T$, as well as the qudit Fourier transform and powers of $Z$ \cite{Barak:07, Tabia2016}.

\section{Examples}
\label{examples}
\subsection{Qubit SICs}

\begin{figure}[t!]
\centering
    \includegraphics[width=0.8\linewidth]{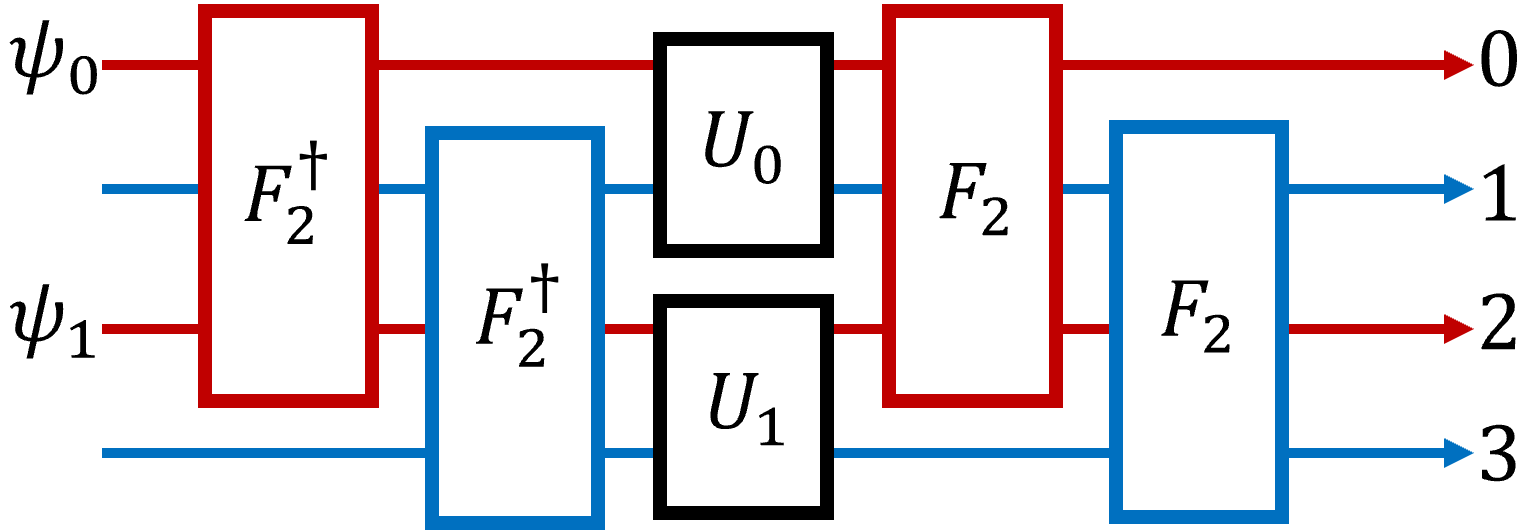}
    \caption{Schematic diagram illustrating the optical implementation of a rank-one
    Weyl--Heisenberg covariant measurement in $d=2$. The input state $|\psi\rangle$ is injected through inputs $(0,2)$. The Fourier matrix $F_2$ and its adjoint $F_2^\dagger$ act on the inputs as indicated by the color coding in the diagram, while the unitary matrices $U_0$ and $U_1$ operate on inputs $(0,1)$ and $(2,3)$, respectively. Each detection event at a numbered output port corresponds to a distinct measurement outcome.}
    \label{fig:qubit}
\end{figure}

We now give a concrete example of this method for constructing a Naimark extension in $d=2$. The input state, its embedding, and the fiducial may be written
\begin{align}
|\psi\rangle = \begin{pmatrix} \psi_0 \\ \psi_1\end{pmatrix} && |\psi, 0\rangle = \begin{pmatrix} \psi_0 \\ 0 \\ \psi_1 \\ 0 \end{pmatrix} && |\phi\rangle = \begin{pmatrix} \phi_0 \\ \phi_1 \end{pmatrix}.
\end{align}
We need to find a unitary matrix of the form 
\begin{align}
    U = \frac{1}{\sqrt{2}}
    \left(\begin{array}{cc|cc}
        \phi_0^* & \cdot & \phi^*_1 & \cdot \\
        \phi^*_0 & \cdot & -\phi^*_1 & \cdot \\
        \hline
        \phi^*_1 & \cdot & \phi^*_0 & \cdot \\
        -\phi^*_1 & \cdot & \phi^*_0 & \cdot
    \end{array}\right),
\end{align}
where dots represent the unknown entries. In order to find them, we first construct a unitary $M$ whose first row is $\langle \phi|$ and whose second row is orthogonal to the first. The natural choice is 
\begin{equation}
    M = 
    \begin{pmatrix}
        \phi^*_0 & \phi^*_1 \\
        -\phi_1 & \phi_0
    \end{pmatrix}.
\end{equation}
Recalling that Fourier matrix in $d=2$ is just the Hadamard matrix $H= \frac{1}{\sqrt{2}}\begin{pmatrix} 1 & 1 \\ 1 & - 1 \end{pmatrix}$, we can construct the blocks $\{S_0, S_1\}$ of the (circulant) matrix $U$ as
\begin{align}
S_0 = \frac{1}{\sqrt{2}}\begin{pmatrix} 1 \\ 1 \end{pmatrix} \begin{pmatrix} \phi^*_0 & - \phi_1 \end{pmatrix} = \begin{pmatrix}\phi^*_0 & - \phi_1  \\ \phi^*_0 & - \phi_1  \end{pmatrix}, &&
S_1 = \frac{1}{\sqrt{2}}\begin{pmatrix} 1 \\ -1 \end{pmatrix} \begin{pmatrix} \phi^*_1 &  \phi_0 \end{pmatrix} = \begin{pmatrix}\phi^*_1 & \phi_0  \\ -\phi^*_1 &  -\phi_0  \end{pmatrix}.
\end{align}
After block diagonalization, the blocks along the diagonal will be $U_j = \sum_k \omega^{-jk} S_k$, or
\begin{equation}
    U_0 = S_0 + S_1 
        = \frac{1}{\sqrt{2}}
        \begin{pmatrix}
        \phi^*_0 + \phi^*_1 & \phi_0 - \phi_1 \\
        \phi^*_0 - \phi^*_1 & -(\phi_0 + \phi_1)
        \end{pmatrix}, 
    \
    U_1 = S_0 - S_1
        = \frac{1}{\sqrt{2}}
        \begin{pmatrix}
        \phi^*_0 - \phi^*_1 & -(\phi_0 + \phi_1) \\
        \phi^*_0 + \phi^*_1 & \phi_0 - \phi_1
    \end{pmatrix}.
\end{equation}
In an optical implementation, the matrices $\{U_0, U_1\}$ act on input modes $(0,1)$ and $(2,3)$, respectively. Together with the Fourier transformation (in this case, easily realized by an unbiased beamsplitter) applied to modes $(0,2)$ and $(1,3)$, we may implement the entire unitary
\begin{align}
    U = \frac{1}{\sqrt{2}}
    \left(\begin{array}{cc|cc}
        \phi_0^* & -\phi_1 & \phi^*_1 & \phi_0 \\
        \phi^*_0 & - \phi_1 & -\phi^*_1 & -\phi_0 \\
        \hline
        \phi^*_1 & \phi_0 & \phi^*_0 & -\phi_1 \\
        -\phi^*_1 & -\phi_0 & \phi^*_0 & - \phi_1
    \end{array}\right),
\end{align}
through its decomposition $U = (F_2^\dagger \otimes I) \text{diag}(U_0, U_1) (F_2 \otimes I)$. A schematic diagram for this style of implementation is depicted in Figure~\ref{fig:qubit}. In particular, we can implement a SIC-POVM by using the fiducial
\begin{equation}
    |\phi\rangle = \frac{1}{\sqrt{6}}
    \begin{pmatrix}
        \sqrt{3+\sqrt{3} }\\
        e^{i\pi/4} \sqrt{3-\sqrt{3}}
    \end{pmatrix}.
\end{equation}
In fact, choosing the second fiducial to have components $(-\phi_1^*, \phi_0^*)$ also gives us a SIC fiducial. Together they form an example of a \emph{compound SIC}, a set of $d$ mutually orthogonal vectors each of whose WH orbit generates a SIC set \cite{SICQuantumKey2020}.

\subsection{Qutrit SICs}

Qutrits have recently been a subject of considerable interest owing to their potential use in reducing circuit depth \cite{ogunkoya2024qutrit}, in quantum optimization \cite{bottrill2023exploring}, and more efficient quantum algorithms \cite{gokhale2020extending}. \cite{Goss2024} considers in particular superconducting qutrit processors. We can implement a qutrit SIC-POVM by using the well known Hesse SIC fiducial \cite{dang2013linear}
\begin{align}
|\phi\rangle = \frac{1}{\sqrt{2}}\begin{pmatrix} 0 \\ 1 \\ -1 \end{pmatrix}.
\end{align}
An appropriate unitary matrix $M$ may be constructed as

\begin{equation}
    M = \frac{1}{\sqrt{2}}
    \begin{pmatrix}
        0 & 1 & -1\\
        \sqrt{2} & 0 & 0\\
        0 & 1 & 1
    \end{pmatrix}.
\end{equation}
In fact, $\frac{1}{\sqrt{2}}(0, 1, 1)^T$ is also a SIC fiducial. The full unitary $U$ constructed according to our scheme is then

\begin{align}
U=\frac{1}{\sqrt{6}}\left(
\begin{array}{ccccccccc}
 0 & \sqrt{2} & 0 & 1 & 0 & 1 & -1 & 0 & 1 \\
 0 & \sqrt{2} & 0 & \omega^{2} & 0 & \omega^{2} & -\omega & 0 & \omega \\
 0 & \sqrt{2} & 0 & \omega & 0 & \omega & -\omega^{2} & 0 & \omega^{2} \\
 -1 & 0 & 1 & 0 & \sqrt{2} & 0 & 1 & 0 & 1 \\
 -\omega & 0 & \omega & 0 & \sqrt{2} & 0 & \omega^{2} & 0 & \omega^{2} \\
 -\omega^{2} & 0 & \omega^{2} & 0 & \sqrt{2} & 0 & \omega & 0 & \omega \\
 1 & 0 & 1 & -1 & 0 & 1 & 0 & \sqrt{2} & 0 \\
 \omega^{2} & 0 & \omega^{2} & -\omega & 0 & \omega & 0 & \sqrt{2} & 0 \\
 \omega & 0 & \omega & -\omega^{2} & 0 & \omega^{2} & 0 & \sqrt{2} & 0 \\
\end{array}
\right),
\end{align}
where $\omega = e^{2\pi i /3}$. This unitary has a structure similar, but distinct from that used by Tabia \cite{Tabia2012}. $U$ may be block diagonalized as
\begin{equation}
    U = (F_3^\dagger \otimes I) \text{diag}(U_0, U_1, U_2) (F_3 \otimes I), 
\end{equation}
where
\begin{align}
U_0 = \frac{1}{\sqrt{6}}
\begin{pmatrix}
0 & \sqrt{2} & 2 \\
-i\sqrt{3} & \sqrt{2} & -1 \\
i\sqrt{3} & \sqrt{2} & -1
\end{pmatrix}, &&
U_1 = \frac{1}{\sqrt{6}}
\begin{pmatrix}
-i\sqrt{3} & \sqrt{2} & -1 \\
i \sqrt{3} & \sqrt{2} & -1 \\
0 & \sqrt{2} & 2
\end{pmatrix}, &&
U_2 = \frac{1}{\sqrt{6}}
\begin{pmatrix}
i \sqrt{3} & \sqrt{2} & -1 \\
0 & \sqrt{2} & 2 \\
-i \sqrt{3} & \sqrt{2} & -1
\end{pmatrix}.
\end{align}
We note that $U_0, U_1, U_2$ are related by permutations of their rows, which may lead to further simplification at an implementation level.

\subsection{Ququart SICs}

Four-dimensional quantum systems offer enhanced capabilities for quantum information processing \cite{seifert2023exploring}, the generation of hyperentanglement \cite{ghosh2021creating}, and hardware-efficient quantum error correction \cite{Brock2025}. Dimension four is also the first dimension for which the general algebraic number-theoretic method behind the construction of SIC-POVMs becomes applicable \cite{bengtsson2017number}. It is also the smallest Hilbert space dimension in which SIC-POVMs can exhibit nontrivial composite structure \cite{Zhu2010Twoqubit}. One of the SIC-POVM fiducial states in $d=4$ is
\begin{align}
    |\psi \rangle = \sqrt{\frac{1-1/\sqrt{5}}{8}}
    \begin{pmatrix}
        e^{-i\pi/4}+1\\
        -i(\alpha e^{-i\pi/4}+1)\\
         e^{-i\pi/4}-1\\
        i(\alpha e^{-i\pi/4}-1)\\
    \end{pmatrix} \quad \text{with} \ 
    \alpha = \sqrt{2+\sqrt{5}},
\end{align}
where the nested square roots are the signature of the number field associated with this SIC. (We note, however, that we could have written the normalization and $\alpha$ in terms of $\sin(\pi/5)$ and $\cos(\pi/5)$, respectively.) $M$ may be chosen to be
\begin{align}
    M  = \sqrt{\frac{1-1/\sqrt{5}}{8}}\left[e^{i\pi/4}
    \begin{pmatrix}
        1 & i\alpha & 1 & -i\alpha\\
        -1 & i & -1 & -i \\
        \alpha  & -i & \alpha  & i \\
        -1 & -i & -1 & i
    \end{pmatrix}
     + 
     \begin{pmatrix}
        1 & i & -1 & i \\
        -\alpha & i & \alpha  & i \\
        -1 & i & 1 & i \\
        1 & i\alpha  & -1 & i\alpha
    \end{pmatrix}
    \right],    
\end{align}
where we have expressed $M$ as the sum of two matrices for clarity. In fact, the rows of $M$ correspond to mutually orthonormal SIC fiducials, providing another example of a compound SIC \cite{SICQuantumKey2020}. The four unitaries in the block decomposition of $U$, explicitly, are
\begin{align}
U_0 = \sqrt{\frac{1-1/\sqrt{5}}{8}}\left[
    e^{i\pi/4}
\left(
\begin{array}{cccc}
 i & i & i & i \alpha  \\
 1 & -\alpha  & -1 & 1 \\
 -i & -i & -i & -i \alpha  \\
 1 & -\alpha  & -1 & 1 \\
\end{array}
\right)
+ 
\left(
\begin{array}{cccc}
 1 & -1 & \alpha  & -1 \\
 \alpha  & 1 & -1 & -1 \\
 1 & -1 & \alpha  & -1 \\
 -\alpha  & -1 & 1 & 1 \\
\end{array}
\right) \right],
\end{align}

\begin{align}
U_1 = \sqrt{\frac{1-1/\sqrt{5}}{8}}\left[
    e^{i\pi/4}
\left(
\begin{array}{cccc}
 1 & -\alpha  & -1 & 1 \\
 -i & -i & -i & -i \alpha  \\
 1 & -\alpha  & -1 & 1 \\
 i & i & i & i \alpha  \\
\end{array}
\right)
+ 
\left(
\begin{array}{cccc}
 \alpha  & 1 & -1 & -1 \\
 1 & -1 & \alpha  & -1 \\
 -\alpha  & -1 & 1 & 1 \\
 1 & -1 & \alpha  & -1 \\
\end{array}
\right) \right],
\end{align}

\begin{align}
U_2 = \sqrt{\frac{1-1/\sqrt{5}}{8}}\left[
    e^{i\pi/4}
\left(
\begin{array}{cccc}
 -i & -i & -i & -i \alpha  \\
 1 & -\alpha  & -1 & 1 \\
 i & i & i & i \alpha  \\
 1 & -\alpha  & -1 & 1 \\
\end{array}
\right)
+ 
\left(
\begin{array}{cccc}
 1 & -1 & \alpha  & -1 \\
 -\alpha  & -1 & 1 & 1 \\
 1 & -1 & \alpha  & -1 \\
 \alpha  & 1 & -1 & -1 \\
\end{array}
\right) \right],
\end{align}

\begin{align}
U_3 = \sqrt{\frac{1-1/\sqrt{5}}{8}}\left[
    e^{i\pi/4}
\left(
\begin{array}{cccc}
 1 & -\alpha  & -1 & 1 \\
 i & i & i & i \alpha  \\
 1 & -\alpha  & -1 & 1 \\
 -i & -i & -i & -i \alpha  \\
\end{array}
\right)
+ 
\left(
\begin{array}{cccc}
 -\alpha  & -1 & 1 & 1 \\
 1 & -1 & \alpha  & -1 \\
 \alpha  & 1 & -1 & -1 \\
 1 & -1 & \alpha  & -1 \\
\end{array}
\right)\right].
\end{align}


\section{Generalized Bell-basis Implementation}
\label{gen_bell_img}

We now take a different, but equivalent point of view on the same Naimark extension. Whereas its block-matrix formulation relies on the three assumptions listed in Section~\ref{from blocks}, we will now see that these assumptions are equivalent to a single assumption: that each choice of embedding $|\psi, i\rangle$ for $i=0,\dots, d-1$ realizes a WH covariant measurement with a different fiducial. Unitarity of the Naimark extension implies that these fiducials must be mutually orthogonal. To see this, we begin with the following observation: the Naimark extension in Section~\ref{from blocks} corresponds to making a generalized Bell-basis measurement on the system and ancilla. 

Recall that the vectorization of a matrix $A$ can be expressed as $|A\rangle = (A\otimes I)\sum_l |l,l\rangle = (I \otimes A^T)\sum_l |l, l\rangle$. The generalized Bell-basis can be defined in terms of the vectorization of the Weyl-Heisenberg operators,
\begin{align}
|D_{jk}\rangle = (D_{jk}\otimes I)\frac{1}{\sqrt{d}}\sum_l |l,l\rangle=\frac{1}{\sqrt{d}}\sum_l \omega^{kl} |j+l, l\rangle.
\end{align}
Since $(A|B) = \tr(A^\dagger B)$ reproduces the Hilbert-Schmidt inner product, $\{|D_{jk}\rangle\}$ forms an orthonormal and, indeed, maximally entangled basis for $\mathcal{H}_d \otimes \mathcal{H}_d$. Suppose we prepare $|\psi\rangle \otimes |\phi^*\rangle$, where $|\psi\rangle$ is the input state and $|\phi\rangle$ is the desired WH fiducial state, and then performing a measurement in the generalized Bell-basis. The amplitudes for the outcomes are
\begin{align}
\langle D_{jk}|(|\psi\rangle \otimes |\phi^*\rangle) &= \frac{1}{\sqrt{d}}\sum_l\langle l, l| (D_{jk}^\dagger \otimes I)(|\psi\rangle \otimes |\phi^*\rangle)\\
&= \frac{1}{\sqrt{d}}\sum_l \langle l|D_{jk}^\dagger |\psi\rangle \langle l|\phi^*\rangle = \frac{1}{\sqrt{d}}\langle \phi |D_{jk}^\dagger |\psi\rangle,
\end{align}
so that the probabilities are precisely those for a WH covariant POVM with fiducial $|\phi\rangle$: $P(j,k) = \frac{1}{d}|\langle \phi_{jk}|\psi\rangle|^2$.

Now let $\mathcal{D} = \sum_{jk} |j,k\rangle\langle D_{jk}|$ be the unitary matrix whose rows are $\langle D_{jk}|$ and which enacts the shift from the computational basis to the generalized Bell-basis. Meanwhile, let $M$ be a unitary matrix whose first row is $\langle \phi|$ so that $M^T|0\rangle = |\phi^*\rangle$ prepares the complex conjugate of the fiducial. Clearly, preparing $|\psi, 0\rangle$, applying these two unitaries, and then measuring in the computational basis implements the POVM. We will now show that the product of the two unitaries,
\begin{align}
U=\mathcal{D}(I \otimes M^T) = \frac{1}{\sqrt{d}}\sum_{jkl}|j,k\rangle\langle l, l| (D_{jk}^\dagger \otimes M^T)=\frac{1}{\sqrt{d}}\sum_{jkl} |j,k\rangle\langle l, l|(M D_{jk}^\dagger \otimes I),
\end{align}
is precisely the same unitary we considered in Section \ref{from blocks}. To see this, we calculate the matrix elements,
\begin{align}
\langle r, s|U|t,u\rangle &= \frac{1}{\sqrt{d}}\sum_{jkl}\langle r,s|j,k\rangle\langle l,l|(MD_{jk}^\dagger \otimes I)|t, u\rangle\\
&= \frac{1}{\sqrt{d}}\sum_{l}\langle l|MD_{rs}^\dagger|t\rangle \langle l|u\rangle
=\frac{1}{\sqrt{d}}\omega^{-s(t-r)}\langle u|M| t-r\rangle.
\label{element}
\end{align}
In tensor product indexing $|j,k\rangle$, $j$ picks out which block, and $k$ the element within that block. A block circulant matrix enjoys the following invariance: sending $r\rightarrow r+n$ and $t\rightarrow t+n$ must leave the matrix invariant. Indeed, Eq.~(\ref{element}) makes this manifest since each element depends only upon the difference between $t$ and $r$. Moreover, consider a single block: it has elements $\langle s |U^{(r,t)}|u\rangle$. In other words, letting $q=t-r$,
\begin{align}
U^{(r,t)}
&=\sum_{su} \frac{1}{\sqrt{d}}\omega^{-sq}\langle u|M|q \rangle  |s\rangle\langle u| =\frac{1}{\sqrt{d}}\sum_s \omega^{-sq}|s\rangle \sum_u \langle q|M^T|u\rangle \langle u| = |\tilde{f}_q\rangle\langle \tilde{m}_q|,
\end{align}
precisely as before.

Whereas in Section \ref{from blocks}, this structure was derived from three assumptions, we can now see that these may be reduced to a single assumption: for each of $d$ choices of embedding $|\psi, i\rangle$, we ought to realize $d$ different WH-POVMs with mutually orthonormal fiducials. Indeed, 
\begin{align}
U|\psi, i\rangle &= \frac{1}{\sqrt{d}}\sum_{jkl}|j,k\rangle\langle l,l|(MD_{jk}^\dagger \otimes I)|\psi, i\rangle =\frac{1}{\sqrt{d}}\sum_{jk}|j,k\rangle\langle m_i|D_{jk}^\dagger|\psi\rangle ,
\end{align}
where $\langle m_i|=\langle i|M$ is the $i$th row of $M$. Thus for each choice of $i$, we realize a rank-one unbiased Weyl--Heisenberg covariant POVM with fiducial $|m_i\rangle$, and since $U$ and therefore $M$ must be unitary, these fiducials must form an orthonormal basis.  We note that this immediately implies that our examples in $d=2$ and $d=4$ implement the entire compound SIC in one experimental set-up. Moreover, we can now see that the structure depicted in Eq.~(\ref{blockU}) is replicated for each fiducial, each offset by $i$ to the right. To see this, recall that the matrix whose rows are $\{\langle r_j|\}$ can be expressed as $R = \sum_j |j\rangle\langle r_j|$. Suppose in particular that $R$ is a $d^2 \times d^2$ matrix, and each row ought to interleave $d$ vectors, each starting at a different offset, a component every $d$ entries: this can be expressed
\begin{align}
R = \sum_{jkl} |j,k\rangle \langle r_{jkl}, l|,
\end{align}
since for any vector $|v\rangle$ and computational basis state $|l\rangle$, $|v\rangle \otimes |l\rangle$ arranges the components of $|v\rangle$ one every $d$ entries, starting from the $l$th entry, the rest being zero. Finally, taking
\begin{align}
\langle r_{jkl}| = \frac{1}{\sqrt{d}}\langle l|MD_{jk}^\dagger,
\end{align}
we recover our unitary $U$: the $l$th fiducial displaced by $D_{j,k}^\dagger$ appears in the $(j,k)$th row, starting at offset $l$, with components placed every $d$ entries. 

In Section \ref{from blocks}, we showed how $U$ may be block diagonalized leading to a decomposition suitable for, e.g., optical implementations where each path is assigned a basis vector in $\mathcal{H}_{d^2}$. Although we expressed the initial state embedding as $|\psi, 0\rangle$, the tensor product there was purely a convenient notation. In contrast, taking the generalized Bell-basis view on the same unitary leads to a decomposition suitable for multiqudit systems, where one treats $\mathcal{H}_{d^2}$ as $\mathcal{H}_d \otimes \mathcal{H}_d$. Indeed, in the block diagonal scheme, it was crucial to embed the input state into $\mathcal{H}_{d^2}$ as $|\psi,0\rangle$ and not $|0, \psi\rangle$, since only the former delivers a block circulant structure. From the point of view of $\mathcal{H}_d\otimes \mathcal{H}_d$, however, whether one chooses the first or the second qudit to be the system or the ancilla is immaterial. In the latter case, as in \cite{Karimipour2002, King2024, mukherjee2025certifyingdimensionalityquantumchannel}, we may realize the transformation to the generalized Bell-basis through a controlled shift followed by a qudit Fourier transform,
\begin{align}
\mathcal{D} &=\frac{1}{\sqrt{d}}\sum_{jk}|j,k\rangle (D_{jk}|
=\frac{1}{\sqrt{d}}\sum_{jkl}|j,k\rangle\langle l,l|(Z^{-k}X^{-j}\otimes I)\\
&=\frac{1}{\sqrt{d}}\sum_{jkl} \omega^{-kl}|j,k\rangle\langle l+j,l|=\frac{1}{\sqrt{d}}\sum_{klm} \omega^{-kl}|m-l,k\rangle\langle m,l|\\
&= \sum_l \left(\sum_m |m-l\rangle\langle m|\right) \otimes \left(\frac{1}{\sqrt{d}}\sum_{k} \omega^{-kl}|k\rangle\langle l|\right)\\
&= \sum_j X^{-j} \otimes \left(\frac{1}{\sqrt{d}}\sum_{kl} \omega^{-kl}|k\rangle\langle l|\right)|j\rangle\langle j|\\
&=(I\otimes F^\dagger )\left(\sum_j X^{-j}\otimes |j\rangle\langle j|\right) \label{shift},
\end{align}
where we let $m=l+j$. Indeed, this is the qudit generalization of the circuit commonly used to implement a Bell-basis measurement: \text{CNOT} followed by a Hadamard. The full unitary then reads
\begin{align}
U = (I\otimes F^\dagger )\left(\sum_j X^{-j}\otimes |j\rangle\langle j|\right)(I \otimes M^T).
\end{align}
In Appendix \ref{qubits}, we report how the qudit controlled shift may be realized via one-and two-qubit operations when $d=2^n$ for $n$ qubits. Finally, in Appendix \ref{ct_duality}, we show how these two pictures of the same unitary can be related by control/target duality.


\section{Discussion and Conclusion}
\label{conclusion}

Informationally complete measurements are highly prized both theoretically and experimentally for their ability to characterize the behavior of quantum systems. In particular, Weyl--Heisenberg covariant POVMs are (generically) informationally complete, and among them SIC-POVMs have pride of place due, e.g., to their optimality for linear quantum state tomography \cite{Scott_2006}. In this work, we have explored a particularly elegant Naimark extension for WH covariant POVMs from two complementary perspectives. On the one hand, extending the work of Tabia \cite{tabia2012experimental} on implementing low dimensional SIC-POVMs, we construct a block circulant Naimark unitary by making three simplifying assumptions about the structure of the blocks. These assumptions mesh more closely with the structure of the WH group than those made in \cite{tabia2012experimental}, and allow us to reduce the problem of determining the entries of the full $d^2\times d^2$ unitary $U$ to the problem of determining a $d \times d$ unitary $M$ whose first row is related to the desired WH fiducial state. Moreover, because this scheme maintains the block circulant structure of $U$ in any dimension, $U$ can always be block diagonalized via the Fourier transform, allowing for a relatively seamless implementation through multiport optical setups, where the required $d^2$ paths can be individually addressed, as well as on certain trapped-ion systems. Indeed, combined with recent advances in photonic path encoding \cite{Luo2015}, the implementation of high-dimensional informationally complete measurements becomes a practical possibility.

On the other hand, we show that the very same Naimark extension can be understood as a generalized Bell-basis measurement \cite{DAriano2004}. This is perhaps the more productive view when considering multiqudit or multiqubit systems, where one treats a $d^2$-dimensional Hilbert space as the tensor product of two $d$-dimensional spaces. The Naimark extension can then be understood as preparing (the complex conjugate of) the WH fiducial state on the ancilla system, followed by a rotation into the generalized Bell-basis. This is a maximally entangled basis for $\mathcal{H}_{d}\otimes \mathcal{H}_d$ whose basis vectors correspond to vectorized WH displacement operators; the basis transformation itself can be realized by a controlled shift followed by an inverse Fourier transform. While this is phrased in the language of qudits, we show how when $d=2^n$ these operations can be broken down into one-and two-qubit gates. Moreover, this point of view makes clear that the three assumptions that guided the block-matrix construction amount to a single assumption, that the same Naimark extension in fact implements $d$ separate WH-POVMs with mutually orthogonal fiducials. Which one is realized depends upon the initial state of the ancilla system: in the optical case, this becomes the choice of which paths are used to encode the initial state. Indeed, this feature of the Naimark extension naturally enables the implementation of so-called compound SICs, consisting of $d$ SICs arising from mutually orthogonal fiducial states. We give concrete examples of the implementation of compound SIC-POVMs in $d=2$ and $d=4$, while in $d=3$ we are able to exhibit an extension that can realize two (but not three) different SICs. 

Above all, it is the exploitation of group symmetry that makes such an elegant Naimark extension possible, and it is the particular choice of the Weyl--Heisenberg group that allows for such flexible decompositions with such a straightforward palette of operators. Certainly, this type of method can be extended to other maximally entangled operator bases beyond the generalized Bell-basis, in particular to nice error bases \cite{Knill1996, Klappenecker2002, Klappenecker2005}. A question for future research is whether a similarly fruitful block-matrix perspective can be given in these cases. Finally, in this paper, we have not overly specialized to any particular hardware implementation: optimally decomposing the elementary unitaries given here for specific contemporary platforms such as photonic chips, trapped-ion qudits, and superconducting circuits remains an important prerequisite for experimental implementation.


\section*{Acknowledgments}
The authors thank Chris Fuchs, Blake Stacey, and Arjun Dhoot for many helpful and stimulating discussions. This research was supported in part by the National Science Foundation through grants NSF-2210495 and OSI-2328774. Sachin Gupta additionally acknowledges support from the University of Massachusetts Boston College of Science and Mathematics Dean’s Doctoral Research Fellowship, funded by Oracle (Project ID R20000000025727).


\bibliography{references.bib}


\appendix

\section{Qubit implementation}
\label{qubits}

The implementation of the qudit Fourier transform in terms of qubit operations is standard \cite{nielsen2010quantum}. Since the clock and shift operators are Fourier conjugates of each other, the remaining question is how to implement the qudit controlled clock operation $CZ$ in terms of one and two qubit gates (assuming that $d=2^n$ for $n$ qubits) \cite{draper2000additionquantumcomputer, RuizPerez2017}. In fact, one can realize this using the same ingredients as the qudit Fourier transform: the Hadamard gate ($\T{H}$), a controlled phase gate $\T{CR}(k)$, and the swap gate $\T{SWAP}$:
\begin{align}
\T{H}= \frac{1}{\sqrt{2}}\begin{pmatrix}1 & 1 \\ 1 & -1\end{pmatrix}	&&\T{R}(k) &= \begin{pmatrix}1 & 0 \\ 0 & e^{2\pi i/2^k} \end{pmatrix}
\end{align}
\begin{align}
\T{CR}(k) &= \begin{pmatrix}I & 0 \\ 0 & R(k)\end{pmatrix} && \T{SWAP} = \begin{pmatrix} 1 & 0 & 0 & 0\\ 0 & 0 & 1 & 0 \\ 0 & 1 & 0 & 0 \\ 0 & 0 & 0 & 1\end{pmatrix}.
\end{align}
Indexing the $n$ qubits by $\{q_j\}_{j=0}^{n-1}$, we can construct the clock operator as
\begin{align}
\T{Z}= \prod_{j=0}^{n-1} \T{R}_{q_j}(j+1) && \T{X}=\T{F}^\dagger \T{Z}\T{F}.
\end{align}
Here $\T{R}_{q_j}$ denotes $\T{R}$ applied to the $q_j$'th qubit, and products ought to be understood from right to left. It is easiest to see that this works by direct calculation, e.g., for two qubits ($d=2^2=4$),
\begin{align}
\T{R}(1)\otimes \T{R}(2)
&= \begin{pmatrix} 1 & 0 \\ 0 & e^{\pi i }\end{pmatrix} \otimes   \begin{pmatrix} 1 & 0 \\ 0 & e^{\pi i/2}\end{pmatrix} 
=\begin{pmatrix}1 & 0 & 0 & 0\\
0 &e^{\pi i /2} & 0 & 0 \\
0 & 0 & e^{\pi i } & 0 \\
0 & 0 & 0 & e^{3\pi i  /2} \end{pmatrix} 
=\sum_m e^{2\pi i m/4}|m\rangle\langle m|=\T{Z}.
\end{align}
Next, we need controlled counterparts of these operators. We first construct a qubit-controlled $\T{Z}$ operator,
\begin{align}
\T{QCZ}_{c,t} &= 	\prod_{j=0}^{n-1} \T{CR}_{c,t_j}(j+1),
\end{align}
which performs $\T{Z}$ on $n$ qubits indexed by $\{t_i\}_{i=1}^n$ conditional on the state of a control qubit $c$. The full qudit-controlled $\T{Z}$ operator, which performs $Z^k$ conditional on the control qudit being in the $|k\rangle$ state, may then be constructed as
\begin{align}
\T{CZ}_{c, t}	=\prod_{j=0}^{n-1} \T{QCZ}^{2^{j}}_{c_{n-j-1, t}},
\end{align}
where the control qudit is realized by $n$ qubits indexed by $\{c_j\}_{j=0}^{n-1}$, and the target qudit is realized by $n$  qubits indexed by $\{t_j\}_{j=0}^{n-1}$.  
Again, the logic is easiest to see by examining a simple case. Let $n=3$, and take the first three qubits to be the target, and the second three to be the control. Denoting $|0\rangle\langle 0| \equiv 0$ and $|1\rangle\langle 1 | \equiv 1$ and suppressing the tensor product sign, so that \!$ZII1=Z \otimes I \otimes I \otimes |1\rangle\langle 1| $, where the first tensor factor is a $2^3=8$ dimensional qudit, and the latter three tensor factors are treated as qubits,
\begin{align}
\T{CZ}_{c,t}&=\T{QCZ}^{2^2}_{0, t}\T{QCZ}^{2^1}_{1, t}\T{QCZ}^{2^0}_{2,t}\\
&=\Big(I0 II + Z^41II\Big)\Big(II0I+Z^2I1I\Big)\Big(III0+ZII1\Big) \\
&=\Big(I00I+Z^201I+Z^410I+Z^611I\Big)\Big(III0 + ZII 1\Big)\\
&=I000+Z001+Z^2010+Z^3011+Z^4100+Z^5101+Z^6110+Z^7111\\
&= \sum_{m=0}^{7}Z^m \otimes |m\rangle\langle m| .
\end{align}
Finally, $\T{CX}$ can be constructed by first applying $\T{F}$ to the target qubits, then $\T{CZ}$, followed by $\T{F}^\dagger$ on the target qubits. 

\section{Control/target duality}
\label{ct_duality}

Returning to the block diagonalization of $U$ in Eq.~(\ref{block_diag}), we may derive a similar expression to Eq.~(\ref{shift}),
\begin{align}
U &= (F^\dagger \otimes I)\left(\sum_j |j\rangle\langle j| \otimes F^\dagger Z^{-j} M^T\right) (F\otimes I)\\
&= \left[(F^\dagger \otimes F^\dagger)\left(\sum_j |j\rangle\langle j| \otimes Z^{-j},\right)(F \otimes I)\right](I \otimes M^T)\label{clock},
\end{align}
from which it follows that we can alternatively realize $\mathcal{D}$  using Fourier transforms and a controlled clock operation. Note that the role of control and target have swapped places as compared to Eq.~(\ref{shift}). In practice one may be more convenient than the other.

It may be clarifying to see how these expressions can be derived directly from each other. Recall that an interaction with Hamiltonian $A\otimes B$ can be interpreted in two alternative ways,
 \begin{align}
 	e^{i(A \otimes B)}=\sum_a|a\rangle\langle a| \otimes e^{iaB}=\sum_b e^{ibA}\otimes |b\rangle\langle b|,
 \end{align}
where $A=\sum_a a|a\rangle\langle a|$, $B=\sum_b b|b\rangle\langle b|$ are spectral decompositions. In other words, the very same interaction can be interpreted as controlled-$B$ operation, conditional on the observable $A$, or as a controlled $A$-operation, conditional on the observable $B$. In the present case, since $Z=\sum_m \omega^{m}|m\rangle\langle m|$, we may define a discrete position operator $Q=\sum_m \frac{2\pi m}{d}|m\rangle\langle m|$ such that $Z = e^{iQ}$. Since $X = F^\dagger Z F$, if we define the discrete momentum operator $P = \sum_m \frac{2 \pi m}{d} F|m\rangle\langle m|F^\dagger$, then we must have $X=e^{-iP}$. Putting this together, 
\begin{align}
\sum_j F^\dagger |j\rangle\langle j| F \otimes Z^{-j} =  e^{i(P \otimes Q)}=\sum_j X^{-j}\otimes |j\rangle\langle j|
\end{align}
from which it is clear why Eq.~(\ref{shift}) is equivalent to Eq.~(\ref{clock}).

\end{document}